%Paper: chao-dyn/9304004
%From: "Shouhong Wang" <wang@beta.math.indiana.edu>
%Date: Tue, 13 Apr 93 21:11:40 EST

\input amstex
\documentstyle{amsppt}
\magnification=\magstep1
\nopagenumbers
\pageno=1

\def\bbR{\Bbb{R}}
\def\bbT{\Bbb{T}}
\def\bbZ{\Bbb{Z}}

\def\sqr#1#2{{\vcenter{\hrule height.#2pt
	\hbox{\vrule width.#2pt height#1pt\kern#1pt
	\vrule width.#2pt}
	\hrule height.#2pt}}}
\def\square{\mathchoice\sqr{5.5}4\sqr{5.0}4\sqr{4.8}3\sqr{4.8}3}
\def\qed{\hskip4pt plus1fill\ $\square$\par\medbreak}

\TagsOnLeft
\hsize 32pc
\vsize 42pc
\NoBlackBoxes
\document

\hfill {To appear in  {\bf J. Funct. Analysis}}
\vskip1.3cm
\bigskip

\topmatter
\title  INERTIAL FORMS OF NAVIER--STOKES EQUATIONS ON THE SPHERE
\endtitle

\endkeywords

\abstract{The Navier--Stokes equations for incompressible flows past a
two--dimensional sphere are considered in this article.  The
existence of an inertial form of the equations is established.
Furthermore for the first time for fluid equations, we derive an
upper bound on the dimension of the differential system (inertial
manifold) which fully reproduces the infinite dimensional dynamics.
This bound is expressed in terms of Grashof Numbers.}
\endabstract
\rightheadtext{Inertial Forms of Navier-Stokes Equations on $S^2$}
\leftheadtext{R. Temam \&  S. Wang}
\author Roger Temam$^{1, 2}$ \&  Shouhong Wang$^1$
\endauthor
\thanks{$^1$ The Institute
for Applied Mathematics \& Scientific Computing,
Indiana University, Bloomington, IN 47405 \endgraf
$^2$ Laboratoire d'Analyse
Num\'erique, Universit\'e  Paris-Sud, Batiment 425, 91405  Orsay,
France}
\endthanks

\endtopmatter

\par

\heading Introduction
\endheading

\bigskip

Inertial manifolds are smooth finite dimensional manifolds that
attract all orbits of a dissipative dynamical system at an
exponential rate.  By restricting the dynamical system to this
manifold we obtain a finite dimensional one called the inertial form
of the system.  The inertial form of a system  has exactly the same
dynamics as the initial one, due in particular to the so--called
asymptotic completeness property (cf. [CFNT], [FSTi]).

The introduction of inertial manifolds in [FST] and their name were
motivated by the hope that inertial manifolds would exist for the
Navier--Stokes equations (at least in dimension two) and that they
would be able to describe the {\it inertial regimes} of turbulence.
This project has been hampered by a restrictive condition for the
existence of inertial manifolds which appeared in [FST] and in all
subsequent works on inertial manifolds, namely the spectral gap
condition.  This condition is not satisfied by the Navier--Stokes
equations, even in space dimension two.  A breakthrough was recently
made by M. Kwak [K1] who embedded the Navier--Stokes equations on a
torus (space periodic flows) into a reaction diffusion system for
which the spectral gap condition is satisfied when the ratio of the
two periods is a rational number.  In this case Kwak obtains,  in this
manner,  an inertial form for the Navier--Stokes equations.
%
%%%%%%%

In this article we want to address the case, very important for
meteorology, of the Navier--Stokes equations past the sphere.  We use
the same methods as in [K1] but our treatment is totally different
because, in particular, we encounter here  the difficulties
related to the geometry.  We
are able as in [K1] to reduce these equations to a reaction--diffusion
system for which the spectral gap condition is satisfied.
Furthermore we are able in our  case to derive an upper bound on  the
dimension of the inertial manifold in terms of the Grashof number.
Although this bound is high, it is still polynomial (and not
exponential) with respect to the Grashof number.

The article is organized as follows.  Section 1 describes the
Navier--Stokes equations on $S^2$.  In Section 2 we introduce the
embedded system and study its  relation to the initial one.  Absorbing
sets and attractors for the embedded system are  studied in Section 3.
Inertial manifolds are obtained in Section 4.  The dimension of the
inertial manifold is estimated in Section 5.  Finally we conclude in
Section 6 with some complementary remarks on flows past a torus and on the
barotropic equation of the atmosphere.

\bigskip
\heading{1.  Navier-Stokes Equations on  $S^2$}
\endheading
\bigskip

Consider the spherical coordinate system $(x^1,x^2) =
(\theta,\varphi)$ on $S^2$, which $\theta\in (0,\pi)$ the colatitude
and $\varphi\in (0,2\pi)$ the longitude.  The Riemannian metric is
given by
$$
(g_{ij}) = \left(\matrix
1 & 0\cr
	0 & \sin^2\theta\cr
\endmatrix\right) ,
$$
with its inverse denoted by $(g^{ij})$.  For geometric notations, see
e.g. [A]  and for, Navier-Stokes flows, Sec. 3.4, Chapt. III in [T1].

For simplicity, we use $u = u^i {\partial\over\partial x^i} = (u^i)$
to denote a vector field on $S^2$.  We can then write the 2D
Navier--Stokes equations on $S^2$ as follows:
$$\left\{\aligned
& u_t + \nabla_u u + \nabla p - \nu\Delta u = f,\\
& \hbox{div}\ u = 0,\\
& u\big|_{t=0} = u_0,\\
\endaligned \right.\tag1.1$$
where the covariant derivative $\nabla_uu$, the divergence div, the
gradient $\nabla$ and the Laplace--Beltrami operator $\Delta$ are
defined by
$$
\left\{\aligned
& \nabla_uu = u^ku^i_{;k}{\displaystyle\partial\over\partial x^i},\\
& \hbox{div}\ u = u^i_{;i},\\
& \nabla p = p_{;k}g^{ik}{\displaystyle\partial\over\partial x^i},\\
& (\Delta u)^i = g^{kl} u^i_{;kl}-u^i.
\endaligned \right. \tag1.2
$$
As we discussed in the introduction, we consider in this article the
stream function formulation of the Navier--Stokes equations.  To this
end, we define the curl operators for both scalar and vector
functions as follows:
$$\left\{ \aligned
&\hbox{curl}\ \psi = -n\times \nabla\psi,\\
& \hbox{curl}\ u = -\hbox{div}\ (n\times u),\\
\endaligned
\right.$$
where $n$ is the outward normal vector, $\psi$ is a scalar function,
and $u$ is a vector function on $S^2$.  Then we can show that
$$\left\{ \aligned
& \nabla_uu = \nabla\left({\displaystyle u^2\over 2}\right) - u\times\,
	\hbox{curl}\ u,\\
& \Delta u = \nabla(\hbox{div}\, u)+ (u\times n)\,\hbox{curl}\ u.
\endaligned \right. \tag1.3$$

Now we set
$$\left\{ \aligned
& u = \hbox{curl}\ \psi,\\
& \zeta = \hbox{curl}\ u = -\Delta\psi,\\
\endaligned\right. \tag1.4$$
where $\psi$ is the stream function.  For simplicity, let
$$
A_0 = -\Delta: H_0\to H_0,\leqno(1.5)
$$
be the minus Laplace--Beltrami operator on $S^2$ for scalar functions
with inverse $A^{-1}_0: H_0\to H_0$.  The function space $H_0$ here
and the function space $V_0$ used later on are defined by
$$\left\{ \aligned
& H_0 = L^2 (S^2)/\bbR = \{\zeta\in L^2 (S^2)|
	{\displaystyle\int_{S^2}}\zeta
	dS^2 = 0\},\\
&V_0 = \dot H^1(S^2),\\
\endaligned \right. \tag1.6
$$
where $\dot H^s(S^2) = H^s(S^2)/\bbR$ for any $s\in\bbR$.  Then (1.4)
can be written as
$$
\zeta = A_0\psi,\ \psi = A^{-1}_0\zeta,\ u =
\hbox{curl}\,(A^{-1}_0\zeta). \leqno(1.7)
$$

We then infer from (1.1) the following stream function formulation of
the 2D Navier--Stokes equations on the sphere:
$$\left\{ \aligned
& \zeta_t + \hbox{div}\, [\zeta u] + \nu A_0\zeta = F =
	\hbox{curl}\, f,\\
& u = \hbox{curl}\, (A^{-1}_0\zeta),\\
& \zeta\big|_{t=0} = \zeta_0.\\
\endaligned \right. \tag1.8
$$

Obviously, the operator $A_0$ is an unbounded, self--adjoint linear
operator with compact inverse, and with domain $D(A_0) = \dot H^2
(S^2)$.  Moreover, using the spectrum of $A_0$, we can define its
power $A^s_0, s\in\bbR$ and
$$
A^s_0 : \dot H^{s+2}\to \dot H^s,\ \forall s\in \bbR
$$
is an isomorphism.
%%%%%

%%%%%
\heading{2.  The Embedded System}
\endheading
\bigskip
\subheading {\bf 2.1.  The Equations for $\nabla\zeta$ and $\zeta u$}

Let $\zeta_0\in D(A_0)$ and $\zeta(t)$ be a solution of (1.5).  We
define an injection map $J : V_0\to H$ by
$$
J(\zeta) = (\zeta,\tilde v,\tilde w)\leqno(2.1)
$$
where
$$\left\{ \aligned
& \tilde v = \nabla \zeta,\\
&\tilde w = \zeta\,\hbox{curl}\, (A^{-1}_0\zeta) = \zeta u.
\endaligned
\right. $$
The Hilbert space $H$ used here and the Hilbert space $V$ which will
be used later, are defined by (cf. Remark 2.1 below)
$$\left\{\aligned &H = H_0 \times H_1\times H_2,\\
& H_1 = \nabla H^1 (S^2), H_2 = L^2 (TS^2),\\
& V = V_0 \times V_1\times V_2,\\
& V_1 = \nabla H^2 (S^2), V_2 = H^1 (TS^2).\\
\endaligned\right.
\leqno(2.2)$$

According to Hodge decomposition theorem (cf. [A]), we have
$$
C^\infty (TS^2) = \{\nabla\phi\ |\ \phi \in C^\infty (S^2)\}\oplus
\{\hbox{curl}\,\psi\ |\ \psi \in C^\infty (S^2)\},
$$
which implies that
$$
H^s(TS^2) = \nabla H^{s+1} (S^2)\oplus \hbox{curl}\, H^{s+1} (S^2),\
\forall s\geq 0.\leqno(2.3)
$$
Therefore, $H_1$ and $V_1$ in (2.2) are well--defined.

Moreover, for simplicity, we need to define two linear unbounded
operators $A_i$ on $H_i$ $(i=1,2)$ as follows:
$$\left\{\aligned &A_1 = -\Delta : H_1\to H_1,\\
& A_2 = -\Delta : H_2 \to H_2,\\
\endaligned \right. $$
the operator $A_1$ being the restriction of $-\Delta$ in $H_1 =
\nabla H^1 (S^2)$ with domain $D(A_1) = \nabla H^3(S^2)$, and the
operator $A_2$ being the minus Laplacian $-\Delta$ in $H_2 = L^2
(TS^2)$ with domain $D(A_2) = H^2 (TS^2)$.

Then we have
\bigskip
{\bf Lemma 2.1.}  {\it The functions $J(\zeta)$ defined by {\rm (2.1)}
satisfies the following system of equations}
$$
\left\{\aligned
&\zeta_t +\,\hbox{div}\,\tilde w + \nu A_0\zeta = F,\\
&\tilde v_t + \nabla\, (\hbox{div}\,\tilde w) + \nu A_1\tilde v= \nabla
	F,\\
&\tilde w_t + \nu A_2\tilde w = F u + \zeta f - 2\nu Tr[\tilde
	v\otimes\nabla u] - (\tilde v\cdot u)u - \zeta\ \hbox{curl}\,
	[A^{-1}_0 (\tilde v\cdot u)],\\
\endaligned \right.
\leqno(2.4)$$
{\it where} $u = \hbox{curl}\, (A^{-1}_0\zeta)$.
\bigskip
{\bf Proof.}  Obviously, (2.4)$_1$ is the same as (1.8).  Applying
the gradient operator to (2.4)$_1$, we obtain easily (2.4)$_2$.

By definition, and using (1.8) and $u_t = \hbox{curl}\,(A^{-1}_0
\zeta_t)$, we find
$$\eqalign{
\tilde w_t &= \zeta_t u + \zeta u_t\cr
%\noalign{\medskip}
&= \{F - \hbox{div}\, [\zeta u] - \nu A_0\zeta\} u +
\zeta\,\hbox{curl}\,\{A^{-1}_0[F - \hbox{div}\,[\zeta u] - \nu
A_0\zeta]\}.\cr}
$$
We then infer from the identity
$$
\Delta(av) = (\Delta a) v + a\Delta v + 2Tr ((\nabla a)\otimes \nabla
v),
$$
that
$$
\Delta\tilde w = (\Delta \zeta) u + \zeta\Delta u + 2Tr
((\Delta\zeta) \otimes \nabla u).
$$
Therefore,
$$\eqalign{
\tilde w_t &- \nu\Delta\tilde w\cr
%\noalign{\medskip}
&= Fu + \zeta f - u\,\hbox{div}\, (\zeta u) -
\zeta\ \hbox{curl}\,[A^{-1}_0 (\hbox{div}\,(\zeta u))] -
2\nu Tr[(\nabla\zeta)\otimes (\nabla u)],\cr
}$$
which implies (2.4)$_3$, using
$\hbox{div}\,(\zeta u) = \zeta\,\hbox{div}\, u + (\nabla \zeta) \cdot
u = \tilde v\cdot u. $
The proof is complete.
\qed
\bigskip
\subheading {\bf 2.2.  The Embeded System}

Equations (2.4) by themselves do not seem to be dissipative.
Therefore in order to capture the dynamics of the NSE (1.1) or
(1.5), we shall modify (2.3) in such a way that the resulting system
satisfies the following properties:

(a) The dynamics on the section $J(V_0)$ is unchanged,

(b) The modified system is dissipative, and

(c) All the solutions of the modified system will converge to the
section $J(V_0)$ as $t\to\infty$.

More previsely, we propose the following modified system of equations:
$$\left\{\aligned
&\zeta_t + \nu A_0\zeta + \hbox{div}\ w = F,\\
& v_t + \nu A_1 v + \nabla (\hbox{div}\ w) = \nabla F,\\
&w_t + \nu A_2 w - Fu - \zeta f + (v\cdot u)u + \zeta\ \hbox{curl}\
	[A^{-1}_0 (v\cdot u)]\\
& \qquad + 2\nu Tr [v\otimes \nabla u] +
	\underline{k\nu [1 + \nu^{-4}\|\zeta\|^4_{H^{1/2}}](w - \tilde
w)} = 0,\\
\endaligned \right. \leqno(2.5)$$
where $\tilde w = \zeta$ curl $(A^{-1}_0\zeta)$.  The system (2.5) is
obtained from (2.4) by adding the underlined term; the positive
constant $k$ will be specified later on.  Generally speaking, $\tilde
v\not= v,\ \tilde w\not= w (\tilde v = \nabla\zeta, \tilde w$ as
before).

For simplicity, we write $Z = (\zeta,v,w)$.  Then we have the
following theorems about the existence and properties of solutions of
the embedded system (2.5).

{\bf Theorem 2.1.}  {\it We assume that $F$ is given in $H_0$, and
$Z_0 = (\zeta_,v_0,w_0)$ is given in $V_0\times H_1\times H_2$.  Then
there exists a unique solution $Z = Z(t)$ of {\rm (2.5)} with initial
value $Z(0) = Z_0$, such that}
$$
Z\in L^2(0,T;D(A_0)\times V_1\times V_2)\cap C([0,T]; V_0\times
H_1\times H_2),\forall\ T > 0.\leqno(2.6)
$$
{\it Moreover, if $f\in V_0$ and $Z_0\in D(A_0)\times V_1\times V_2$,
then}
$$
\left\{\aligned
&Z\in L^2 (0,T;D(A^{3/2}_0)\times D(A_1)\times D(A_2))\cap
	C([0,T]; D(A_0)\times V_1\times V_2),\\
&Z_t\in L^2 (0,T;V_1\times H_1\times H_2).\\
\endaligned\right.
\leqno(2.7)$$
\bigskip
{\bf Theorem 2.2.}  {\it Let $\zeta (t)$ be a solution of
{\rm (1.8)} with initial value $\zeta (0) = \zeta_0 \in D(A_0)$, and
let $f\in V_0$.  Then $Z(t) = J(\zeta (t))$ is a solution of
{\rm (2.5)} with initial value $Z(0) = J(z_0)$ for all $t > 0$.
Conversely, if $Z(t) = (\zeta (t), v(t),w(t))$ is a solution of
{\rm (2.5)} with $Z(0)= J(\zeta_0)\in D(A_0)\times V_1\times V_2$,
then $\zeta (t)$ is a solution of {\rm (1.8)} for $t\geq 0$ and
$Z(t) = J(\zeta(t))$.  In particular,
$J(D(A_0)\times V_1\times V_2)$ is a positively invariant set for
{\rm (2.5)}}.
\bigskip
{\bf Theorem 2.3.}  {\it For any $Z_0 = (\zeta_0,v_0,w_0)\in
V_0\times H_1\times H_2$,
$$
|v(t) - \tilde v(t)|^2 + \nu^{-2}|w(t) - \tilde w(t)|^2\leq
e^{-\nu t}(|v_0 - \tilde v_0|^2 + \nu^{-2}|w_0 - \tilde w_0|^2),\
\forall\ t > 0.\leqno(2.8)
$$
Consequently,}
$$
\lim_{t\to\infty}|Z(t) - J(\zeta (t))| = 0.
$$

The proof of Theorem 2.2 is easy.  We only have to prove Theorems 2.1
and 2.3.
\bigskip

{\bf Proof of Theorem 2.1.}  First of all, we solve the equations
(2.5)$_1$ and (2.9)-(2.10) below for the unknown functions
$(\zeta,v-\tilde v, w - \tilde w)$, by Galerkin method.  Then we
obtain the solution $Z = (\zeta,v,w)$ of (2.5).  Since the Galerkin
method is a standard procedure, here we only present some formal {\it
a priori} estimates, which will also be used later for the existence
of attractors of the embedded system (2.5).

{\bf Step 1.  Equations for} $\zeta, v - \tilde v$ {\bf and} $w -
\tilde w$.  We infer from (2.5)$_1$ that
$$
\tilde v_t + \nu A_1\tilde v + \nabla (\hbox{div}\ w) = \nabla F,
$$
which implies that
$$
(v - \tilde v)_t + \nu A_1(v - \tilde v)= 0.\leqno(2.9)
$$

We also have, as in the proof of Lemma 2.1 and since $\hbox{div}\
\tilde w = \tilde v\cdot u$,
$$
\tilde w_t + \nu A_2\tilde w - F u - \zeta f - 2\nu Tr[\tilde
v\otimes \nabla u] - (\hbox{div}\ w) u - \zeta\ \hbox{curl}\
[A^{-1}_0 (\hbox{div}\ w)] = 0.
$$

Therefore
$$\leqalignno{
(w - \tilde w&)_t +  \nu A_2 (w - \tilde w) &(2.10)\cr
%\noalign{\medskip}
& + (v\cdot u - \hbox{div}\ w)u + \zeta\ \hbox{curl}\
	[A^{-1}_0(v\cdot u - \hbox{div}\ w)]\cr
%\noalign{\medskip}
& + 2\nu Tr [(v - \tilde v)\otimes \nabla u] + k\nu [1 +
	\nu^{-4}\|\zeta\|^4_{H^{1/2}}] (w-\tilde w) = 0.\cr
}$$

{\bf Step 2.  Energy estimates I.}  We infer from (2.5)$_1$ that
$$\eqalign{
{1\over 2} {d\over dt} |\zeta|^2 + \nu|A^{1/2}_0\zeta|^2 &=
	(F,\zeta) - (\hbox{div}\ w,\zeta)\cr
&\leq |F|\cdot |\zeta| + |(\hbox{div}\ (w - \tilde w),\zeta)| +
	 |(\hbox{div}\ \tilde w, \zeta)|\cr
%%\noalign{\medskip}
&\leq |F|\cdot|\zeta| + |w - \tilde w|\cdot |A^{1/2}_0\zeta|\cr
%%\noalign{\medskip}
&\leq {c\over \nu} (|F|^2 + |w-\tilde w|^2) + {\nu\over 2}
|A^{1/2}_0\zeta|^2,\cr
}$$
i.e.
$$
{d\over dt} |\zeta|^2 + \nu|A^{1/2}_0\zeta|^2 \leq {c\over\nu} (|F|^2
+ |w - \tilde w|^2).\leqno(2.11)
$$

By (2.9), we have
$$
{d\over dt} |v - \tilde v|^2 + 2\nu|A^{1/2}_1 (v-\tilde v)|^2=0.\leqno(2.12)
$$

Moreover, we deduce from (2.10) that
$$\leqalignno{
&\ {\alpha\over 2} {d\over dt}|w - \tilde w|^2 + \nu\alpha|A^{1/2}_2
	(w-\tilde w)|^2 + k\nu\alpha [1 + \nu^{-4}
	\|\zeta\|^4_{H^{1/2}}]\cdot |w - \tilde w|^2 &(2.13)\cr
%\noalign{\medskip}
&= -\alpha ((v\cdot u- \hbox{div}\ w)u + \zeta\ \hbox{curl}\
	[A^{1/2}_0 (v\cdot u - \hbox{div}\ w)]\cr
%\noalign{\medskip}
&\qquad + 2\nu Tr [(v - \tilde v)\otimes \nabla u], w-\tilde w),\cr
}$$
where we introduce for convenience a constant $\alpha = \nu^{-2} >
0$.

The $1^{st}$ term in the right hand-side of (2.13) is equal to
$$\eqalign{
-&\alpha([(v-\tilde v)\cdot u + \hbox{div}\ (\tilde w - w)]u,w-\tilde
	w)\cr
%\noalign{\medskip}
&\leq c\alpha \|u\|^2_{L^8}\cdot |w-\tilde w|\cdot | A^{1/2}_1 (v -
	\tilde v)| + c\alpha\|u\|_{L^\infty}\cdot |w-\tilde w|\cdot
	|A^{1/2}_2 (w - \tilde w)|\cr
%\noalign{\medskip}
&\leq (\hbox{since}\ \|\zeta\|_{H^s}\sim
	\|\Delta\psi\|_{H^s}\sim\|\psi\|_{H^{s+2}}\sim
\|\nabla \psi\|_{H^{s+1}})\cr
%\noalign{\medskip}
&\leq {c\over \nu}(\alpha^2 |\zeta|^4 +
	\alpha\|\zeta\|^2_{H^{1/2}})\cdot
	|w - \tilde w|^2 + {\nu\over 4}| A^{1/2}_1 (v - \tilde v)|^2
	+ {\nu\alpha\over 4}|A^{1/2}_2 (w - \tilde w)|^2.\cr
}$$

The $2^{nd}$ term in the right hand side of (2.13) is equal to
$$\eqalign{
- &\alpha (\zeta\,\hbox{curl}\ \{A^{-1}_0[(v-\tilde v)\cdot u +
	\hbox{div}\ (\tilde w - w)]\}, w - \tilde w)\cr
%\noalign{\medskip}
&\leq c\alpha[\|\zeta\|^2_{H^{1/2}}|A^{1/2}_1(v - \tilde v)| +
	\|\zeta\|_{H^{1/2}}|A^{1/2}_2 (w - \tilde w)|]\cdot |w -
	\tilde w|.\cr
}$$

The $3^{rd}$ term in the right hand side of (2.13) is majorized by
$$
c\nu\alpha|w-\tilde w|\cdot\|\zeta\|_{H^{1/2}}|A^{1/2}_1 (v-\tilde
v)|.
$$

Therefore
$$\leqalignno{
\alpha {d\over dt}|w-\tilde w &|^2 + \nu\alpha|A^{1/2}_2 (w-\tilde
	w)|^2 + 2k\nu^{-1}[1 + \nu^{-4}\|\zeta\|^4_{H^{1/2}}]\cdot
	|w-\tilde w|^2&(2.14)\cr
%\noalign{\medskip}
&\leq c\nu^{-1}(1 + \nu^{-4}\|\zeta\|^4_{H^{1/2}})\cdot |w-\tilde
w|^2 + \nu|A^{1/2}_1 (v-\tilde v)|^2.\cr
}$$
Then, with the constant $\alpha = \nu^{-2}$ still at our disposal, we
have
$$\eqalign{
{d\over dt}& [|\zeta|^2 + |v-\tilde v|^2 + \alpha|w - \tilde w|^2]
	+ \nu[|A^{1/2}_0\zeta|^2 + |A^{1/2}_1(v-\tilde v)|^2 +
	\alpha|A^{1/2}_2(w - \tilde w)|^2]\cr
%\noalign{\medskip}
&\qquad + 2k\nu^{-1}[1 + \nu^{-4}\|\zeta\|^4_{H^{1/2}}]\cdot
	|w-\tilde w|^2\cr
%\noalign{\medskip}
&\leq c\nu^{-1}(1 + \nu^{-1}\|\zeta\|^4_{H^{1/2}})\cdot
	|w-\tilde w|^2 + {c\over \nu}|F|^2.\cr
}$$
Hence for $k$ (absolute constant) large enough, we obtain
$$\leqalignno{
{d\over dt}& [|\zeta|^2 + |v-\tilde v|^2 + \alpha |w - \tilde w|^2]
	+ \nu[|A^{1/2}_0\zeta|^2 + |A^{1/2}_1(v-\tilde v)|^2&(2.15)\cr
%\noalign{\medskip}
&\qquad	+\alpha|A^{1/2}_2(w - \tilde w)|^2] + k\nu^{-1}
	[1 + \nu^{-4}\|\zeta\|^4_{H^{1/2}}]\cdot
	|w-\tilde w|^2\cr
%\noalign{\medskip}
&\leq {c\over \nu}|F|^2.\cr
}$$
Furthermore we consider the inner product between (2.5)$_1$ and
$A_0\zeta$, and we obtain
$$\eqalign{
{1\over 2} {d\over dt} |A^{1/2}_0 \zeta|^2 + \nu|A_0\zeta|^2 &=
	(F,A_0\zeta) - (\hbox{div}\ w, A_0\zeta)\cr
%\noalign{\medskip}
&\leq |F|\cdot |A_0\zeta| + |(\hbox{div}\ (w-\tilde w),
	A_0\zeta)| + |(\hbox{div}\ \tilde w,A_0\zeta)|\cr
%\noalign{\medskip}
&\leq {c\over \nu}|F|^2 + {c_1\over 2\nu}|A^{1/2}_2
	(w - \tilde w)|^2 + {\nu\over 4}|A_0\zeta|^2 +
	|\hbox{div}\ \tilde w|\cdot|A_0\zeta|.\cr
}$$
Since
$$
|\hbox{div}\ \tilde w|\cdot|A_0\zeta|\leq {c\over\nu^3}|\zeta|^4\cdot
|A^{1/2}_0\zeta|^2 + {\nu\over 4}|A_0\zeta|^2,
$$
we obtain
$$
{d\over dt}|A^{1/2}_0\zeta|^2 + \nu|A_0\zeta|^2\leq {c\over\nu}|F|^2
+ {c_1\over\nu}|A^{1/2}_2 (w - \tilde w)|^2 + {c\over
\nu^3}|\zeta|^4\cdot |A^{1/2}_0\zeta|^2.\leqno(2.16)
$$
Finally the combination of (2.12), (2.14) and (2.16) shows that
$$\eqalign{
{d\over dt}&\left[{1\over 2c_1}|A^{1/2}_0\zeta|^2 + |v-\tilde v|^2 +
\alpha | w - \tilde w|^2\right]\cr
%\noalign{\medskip}
&+ \nu\left[{1\over 2c_1}|A_0\zeta|^2 + |A^{1/2}_1 (v - \tilde v)|^2
	+ {\alpha\over 2}|A^{1/2}_2 (w - \tilde w)|^2\right]\cr
%\noalign{\medskip}
&\qquad	+ 2k\nu^{-1}[1 + \nu^{-4}\|\zeta\|^4_{H^{1/2}}]\cdot
	|w - \tilde w|^2\cr
%\noalign{\medskip}
&\leq c\nu^{-1}(1 + \nu^{-4}\|\zeta\|^4_{H^{1/2}})\cdot |w-\tilde
	w|^2 + {c\over \nu^3}|\zeta|^4 \cdot |A^{1/2}_0\zeta|^2 +
	{c\over \nu}|F|^2,\cr
}$$
i.e.,
$$\leqalignno{
{d\over dt}&\left[{1\over 2c_1}|A^{1/2}_0\zeta|^2 + |v-\tilde v|^2 +
	\alpha |w - \tilde w|^2\right]&(2.17)\cr
%\noalign{\medskip}
&+ \nu\left[{1\over 2c_1}|A_0\zeta|^2 + |A^{1/2}_1 (v - \tilde v)|^2
	+ {\alpha\over 2}|A^{1/2}_2 (w - \tilde w)|^2\right]\cr
%\noalign{\medskip}
&+ k\nu^{-1}[1 + \nu^{-4}\|\zeta\|^4_{H^{1/2}}]\cdot
	|w - \tilde w|^2\cr
%\noalign{\medskip}
&\leq {c\over\nu^3} |\zeta|^4\cdot |A^{1/2}_0\zeta|^2 + {c\over\nu}
	|F|^2.\cr
}$$

By Gronwall's inequality, we obtain from (2.15) and (2.17) that
$$
(\zeta,v-\tilde v,w-\tilde w)\in L^2 (0,T;D(A_0)\times V_1\times V_2)
\cap L^\infty (0,T;V_0\times H_1\times H_2),\ \forall\
T>0.\leqno(2.18)
$$
Then from equation (2.5)$_1$ and (2.9)--(2.10), we can obtain {\it a
priori} estimates for
$$
(\zeta,v-\tilde v, w-\tilde w)_t \in L^2 (0,T;H_0\times V'_1\times
V'_2),\ \forall\ T>0.
$$
Hence
$$
(\zeta,v-\tilde v, w-\tilde w) \in C([0,T];V_0\times H_1\times H_2),\
\forall\ T>0.\leqno(2.19)
$$

On the other hand, by definition, we can prove that
$$
\left\{\aligned &
|\tilde v| = |\nabla\zeta|\leq |A^{1/2}_0\zeta|,\\
& |A^{1/2}_1\tilde v|\leq|A_0\zeta|,\\
& |\tilde w| = |\zeta\ \hbox{curl}\ (A^{-1}_0\zeta)|\leq
	|A^{1/2}_0\zeta |^2,\\
&|A^{1/2}_2\tilde w|\leq |A_0\zeta|\cdot |A^{1/2}_0\zeta|.
\endaligned\right. \leqno(2.20)
$$
Then the combination of (2.18)--(2.20) proves the existence of a
solution $Z = (\eta,v,w)$ of (2.5) satisfying (2.6).

{\bf Step 3.  Energy estimates II.}  From (2.5)$_1$, we also have
$$\eqalign{
{1\over 2}{d\over dt} |A_0\zeta|^2 + \nu|A^{3/2}_0\zeta|^2 &=
	(F,A^2_0\zeta) - (\hbox{div}\ w,A^2_0\zeta)\cr
%\noalign{\medskip}
&\leq |A^{1/2}_0F|\cdot |A^{3/2}_0\zeta| + |(\hbox{div}\ (w-\tilde w),
	A^2_0\zeta)| + |(\hbox{div}\ \tilde w,A^2_0\zeta)|\cr
%\noalign{\medskip}
&\leq \{|A^{1/2}_0F| + |\nabla(\hbox{div}\ (w-\tilde w))| +
	|\nabla(\nabla\zeta\cdot \hbox{curl}\ (A^{-1}_0\zeta))|\}
	\cdot |A^{3/2}_0\zeta|\cr
%\noalign{\medskip}
&\leq {c\over \nu} |A^{1/2}_0F|^2 +  {c_2\over 2\nu}|A_2
	(w-\tilde w)|^2 + {c\over\nu}\|\zeta\|^2_{H^{1/2}}
	\cdot |A_0\zeta |^2 + {\nu\over 2}|A^{3/2}_0\zeta|^2,\cr
}$$
i.e.
$$
{d\over dt} |A_0\zeta|^2 + \nu|A^{3/2}_0\zeta|^2 \leq
{c\over \nu} |A^{1/2}_0F|^2 + {c_2\over \nu}|A_2 (w-\tilde w)|^2 +
{c\over\nu}\|\zeta\|^2_{H^{1/2}}\cdot|A_0\zeta|^2. \tag2.21
$$

By (2.9), we have
$$
{d\over dt} |A^{1/2}_1 (v-\tilde v)|^2 + 2\nu|A_1(v-\tilde v)|^2
=0.\leqno(2.22)
$$

Moreover from (2.10) we obtain that
$$\leqalignno{&&(2.23)\cr
{\alpha\over 2}& {d\over dt}|A^{1/2}_2 (w-\tilde w)|^2 + \nu\alpha|A_2
	(w-\tilde w)|^2 + k\nu^{-1}[1 +\nu^{-4}\|\zeta\|^4_{H^{1/2}}]
	\cdot|A^{1/2}_2 (w-\tilde w)|^2\cr
%\noalign{\medskip}
&= -\alpha((v \cdot u-\hbox{div}\ w)u +\zeta\  \hbox{curl}\ [A^{-1}_0(v\cdot
	u-\hbox{div}\ w)]\cr
%\noalign{\medskip}
&\qquad +2\nu Tr[(v-\tilde v)\otimes\nabla u], A_2(w-\tilde w)).\cr
}$$
the $1^{st}$ term in the right hand side of (2.23) is equal to
$$\eqalign{
&-\alpha([(v-\tilde v)\cdot u + \hbox{div}\ (\tilde w-w)]u, A_2(w
	-\tilde w))\cr
%\noalign{\medskip}
&\leq c\alpha \{|\zeta|^2 \cdot |A^{1/2}_1 (v-\tilde v)| +
	\|\zeta\|_{H^{1/2}}\cdot |A^{1/2}_2 (w-\tilde w)|\}\cdot|A_2
	(w-\tilde w)|.\cr
}$$
The $2^{nd}$ term in the right hand side of (2.23) is equal to
$$\eqalign{
&-\alpha(\zeta\ \hbox{curl}\ \{A^{-1}_0 [(v-\tilde v)\cdot u
	+ \hbox{div}\ (\tilde w-w)]\}, A_2 (w-\tilde w))\cr
%\noalign{\medskip}
&\leq c\alpha [\|\zeta\|_{H^{1/2}}|\zeta|\cdot |A^{1/2}_1
	(v-\tilde v)| + \|\zeta\|_{H^{1/2}}|w-\tilde w|]\cdot
	|A_2 (w-\tilde w)|.\cr
}$$
The $3^{rd}$ term in the right hand side of (2.23) is majorized by
$$
c\nu^{-1}|A_2 (w-\tilde w)|\cdot\|\zeta\|_{H^{1/2}}|A^{1/2}_1(v-\tilde
v)|.
$$
Therefore
$$\leqalignno{&&(2.24)\cr
\alpha & {d\over dt} |A^{1/2}_2 (w-\tilde w)|^2 + \nu\alpha |A_2
	(w-\tilde w)|^2 + k\nu^{-1}[1 +\nu^{-4}\|\zeta\|^4_{H^{1/2}}]
	\cdot |A^{1/2}_2 (w-\tilde w)|^2\cr
%\noalign{\medskip}
&\leq {c\alpha\over\nu} (|\zeta|^4 + \|\zeta\|^2_{H^{1/2}}\cdot
	|\zeta|^2 + \nu^2\|\zeta\|^2_{H^{1/2}})\cdot |A^{1/2}_1
	(v-\tilde v)|^2\cr
%\noalign{\medskip}
&\qquad + {c\alpha\over\nu} \|\zeta\|^2_{H^{1/2}}|A^{1/2}_2 (w-\tilde
	w)|^2.\cr
}$$

Finally (2.21)--(2.22) and (2.24) imply that
$$\leqalignno{
{d\over dt}& \left[{1\over 2c_2} |A_0\zeta|^2 + |A^{1/2}_1 (v-\tilde
	v)|^2 + \alpha |A^{1/2}_2 (w-\tilde w)|^2\right]&(2.25)\cr
%\noalign{\medskip}
&+ \nu\left[{1\over 2c_2}|A^{3/2}_0\zeta|^2 + 2|A_1(v-\tilde v)|^2 +
	{\alpha\over 2}|A_2(w-\tilde w)|^2\right]\cr
%\noalign{\medskip}
&+ 2 k\nu^{-1}[1 + \nu^{-4}\|\zeta\|^4_{H^{1/2}}]\cdot
	|A^{1/2}_2(w-\tilde w)|^2\cr
%\noalign{\medskip}
&\leq {c\over\nu} |A^{1/2}_0F|^2 + {c\over\nu} [\alpha|\zeta|^4 +
	\alpha\|\zeta\|^2_{H^{1/2}}\cdot|\zeta|^2 +
	\|\zeta\|^2_{H^{1/2}}]\cr
%\noalign{\medskip}
&\qquad \cdot \left[{1\over 2c_2}|A_0\zeta|^2 + |A^{1/2}_1(v-\tilde
	v)|^2 + \alpha |A^{1/2}_2(w-\tilde w)|^2\right].\cr
}$$

Using Gronwall's inequality as in Step 2, we can prove that if
$Z_0\in D(A_0)\times V_1\times V_2$, then $Z$ satisfies (2.7).

The proof of the uniqueness of solution is easy, and is omitted.

Theorem 2.1 is proved.
\qed
\bigskip
\noindent {\bf Proof of Theorem 2.3}.  We infer from (2.12) and
(2.14) that
$$\eqalign{
{d\over dt}[|v-\tilde v|^2 + \alpha|w-\tilde w|^2]&\leq -\nu
	[|A^{1/2}_1(v-\tilde v)^2 + \alpha|A^{1/2}_2 (w-\tilde
	w)|^2]\cr
&\leq -\nu[|v-\tilde v|^2 + \alpha|w-\tilde w|^2],\cr
}$$
which proves (2.8).
\qed
\bigskip
\heading{3.  Absorbing Sets  and Attractors}
\endheading
\bigskip

First of all, Theorem 2.1 indicates that we can define the operators
$$
\Sigma (t) : Z_0\longrightarrow Z(t),\ \ \forall\ t\geq 0.\leqno(3.1)
$$
These operators enjoy the standard semigroup properties, and they are
continuous operators in $V_0\times H_1\times H_2$ and in
$D(A_0)\times V_1\times V_2$.

We now prove the existence of an absorbing set of the semigroup
$\{\Sigma (t)\}_{t\geq 0}$ in $V_0\times H_1\times H_2$.  T othis end, by
Gronwall's inequality, we infer from (2.15) that
$$
|\zeta(t)|^2 + |(v-\tilde v)(t)|^2 + \alpha|(w-\tilde w)(t)|^2\leq
c|F|^2 + e^{-\nu t}(|\zeta_0|^2 + |v_0-\tilde v_0|^2 + \alpha|w_0
-\tilde w_0|^2),
$$
for any $t\geq 0$.  Then for any $R >0$, if $Z_0 =
(\zeta_0,v_0,w_0)\in B_{V_0\times H_1\times H_2}(0,R)$, the ball in
$V_0\times H_1\times H_2$ centered at 0 with radius $R$, there exists
$t_0 = t_0(R) >0$ such that for any $t\geq t_0$
$$\left\{\aligned &
|\zeta(t)|^2 + |(v-\tilde v)(t)|^2 + \alpha|(w-\tilde w)(t)|^2\leq
	c|F|^2,\\
& {\displaystyle\int^{t+1}_t} [|A^{1/2}_0\zeta (t)|^2
	+ |A^{1/2}_1(v-\tilde v)(t)|^2 +
	\alpha|A^{1/2}_2(w-\tilde w)(t)|^2]dt\leq c|F|^2.\\
\endaligned
\right. \leqno(3.2)
$$

Now we return to (2.17).  By the uniform Gronwall lemma (cf. Lemma
1.1 in  [T1]), we obtain that
$$
|A^{1/2}_0\zeta (t)|^2 + |(v-\tilde v)(t)|^2 + |(w-\tilde w)(t)|^2
\leq c|F|^2\exp (c|F|^4),\ \forall\ t\geq t_0 + 1.\leqno(3.3)
$$

By (2.20), (3.3) shows that
$$
|A^{1/2}_0\zeta (t)|^2 + |v(t)|^2 + |w(t)|^2
\leq c|f|^2\exp (c|F|^4) + c|F|^4\exp(c|F|^4),\
\forall\ t\geq t_0 + 1.\leqno(3.4)
$$

Now we define the right hand side of (3.4) as $\rho^2_0$, then we
conclude that
$$\Sigma (t) B_{V_0\times H_1\times H_2}(0,R)\subseteq B_{V_0\times
H_1\times H_2} (0,\rho_0),\ \forall\ t\geq t_0 + 1.\leqno(3.5)
$$
In other words, $B_{V_0\times H_1\times H_2}(0,\rho_0)$ is an
absorbing set in $V_0\times H_1\times H_2$ for the semigroup
$\Sigma(t)$.

Moreover, from (2.17), we also have
$$\leqalignno{
{\displaystyle\int^{t+1}_t}[|A_0\zeta &(t)|^2 + |A^{1/2}_1(v-\tilde
	v) (t)|^2 + \alpha
	|A^{1/2}_2(w-\tilde w) (t)|^2]dt&(3.6)\cr
&\qquad\leq c|F|^2 + c\rho^2_0 (1 + |F|^4),\ \forall\ t\geq t_0 +
1.\cr }
$$

Now we want to establish the existence of an absorbing set in
$D(A_0)\times V_1\times V_2$.  We apply the uniform Gronwall lemma to
(2.25).  Then we obtain
$$
|A_0\zeta (t)|^2 + |A^{1/2}_1(v-\tilde v) (t)|^2 + \alpha
	|A^{1/2}_2(w-\tilde w) (t)|^2\leq c_1 (|A^{1/2}_0 F|),\
\forall\ t\geq t_0 + 2,\leqno(3.7)
$$
where $c_1(|A^{1/2}_0F|)$ is a function of $|A^{1/2}_0F|$.  On the
other hand, it is obvious that
$$\left\{\aligned &
|A^{1/2}_1\tilde v|\leq c|A_0\zeta|,\\
&|A^{1/2}_2\tilde w|\leq c|A_0\zeta|^2.\\
\endaligned\right.
\leqno(3.8)
$$
Therefore
$$
|A_0\zeta (t)|^2 + |A^{1/2}_1v(t)|^2 + \alpha|A^{1/2}_2 w(t)|^2\leq
c_2(|A^{1/2}_0F|),\ \forall\ t\geq t_0 + 2.\leqno(3.9)
$$
where $c_2(|A^{1/2}_0F|)$ is a function of $|A^{1/2}_0F|$.  This
shows that the ball $B_{D(A_0)\times V_1\times V_2}(0,\rho_1)$ in
$D(A_0)\times V_1\times V_2$ is an absorbing set in $D(A_0)\times
V_1\times V_2$, and that the operators $\Sigma (t)$ are uniformly
compact for $t$ large.  Here $\rho^2_1 = c_2(|A^{1/2}_0F|)$.  Hence
by Theorem 1.1, Chapter I of [T1] we promptly obtain that
\bigskip

{\bf Theorem 3.1}.  {\it There exists a unique global attractor
${\Cal A}\subseteq D(A_0)\times V_1\times V_2$ of the embedded system
{\rm (2.5)}, such that ${\Cal A}$ is compact, connected and maximal
in $V_0\times H_1\times H_2$ and $D(A_0)\times V_1\times V_2$.}

The following theorem shows that ${\Cal A}$ is exactly the lifting of
the attractor ${\Cal A}_0$ of the original Navier--Stokes equations
(1.1) or (1.5), via the injection map $J : V_0\to V_0\times H_1\times
H_2$ defined by (2.2).

{\bf Theorem 3.2.}  {\it We have}
$$
{\Cal A} = J(\Cal A_0).\leqno(3.10)
$$
\vskip.5truein
\heading{4.  Inertial Manifolds}
\endheading
\bigskip
\noindent
\subheading{\bf 4.1.  Preliminaries}

The embedded system can be rewritten in the following concise form
$$
{dZ\over dt} + AZ + R(Z) = 0,\leqno(4.1)
$$
where the linear and nonlinear operators $A,R$ are defined by
$$
AZ = \left(\matrix
      \nu A_0\zeta + \hbox{div}\ w\cr
	\nu A_1 v + \nabla(\hbox{div}\ w)\cr
	\nu A_2w\endmatrix\right),\ \
RZ = \left(\matrix
R_0(Z)\cr
	R_1(Z)\cr
	R_2(Z)
\endmatrix\right),\leqno(4.2)
$$
where $R_0(Z) = -F, R_1(Z) = -\nabla F$ and
$$
R_2(Z) = -F u - \zeta f + (v\cdot u)u + \zeta\ \hbox{curl}\
[A^{-1}_0(v\cdot u)] + 2\nu Tr[v\otimes \nabla u] + k\nu
[1 + \nu^{-4}\|\zeta\|^4_{H^{1/2}}](w-\tilde w).
$$

Then we can obtain easily that $D(A) = D(A_0)\times D(A_1)\times
D(A_2)$, and $D(A)$ is dense in $H$.

For technical reasons, we need to use the following equivalent norm
$|\cdot|_H$ in $H$:
$$
|Z|_H \dot= (|\zeta|^2 + |v|^2 + \alpha |w|^2)^{1/2},
$$
the corresponding inner product being denoted by $(\cdot,\cdot)_H$;
here $\alpha = 1/\nu^2$.  Then we have the following Garding
inequality
$$
(AZ,Z)_H\geq {\nu\over 6}\{|A^{1/2}_0\zeta|^2 + |A^{1/2}_1 v|^2 +
\alpha |A^{1/2}_2 w|^2\},\leqno(4.3)
$$
which can be proved easily by taking the following inequality into
consideration:
$$
(\nabla(\hbox{div}\ w),v) = (w,\nabla (\hbox{div}\ v)) = (w,A_1,v)\leq
{1\over 2}|A^{1/2}_1 v|^2 + {1\over 2}|A^{1/2}_2 w|^2.
$$

Moreover, as in [K1], we can prove that $A$ is a sectorial operator,
and $A^{-1}:H\to H$ is a compact linear operator.  Therefore the
spectrum $\sigma(A)$ consists of a countable number of eigenvalues with
no finite accumulation points and each eigenvalue has finite multiplicity.
Indeed we have

{\bf Lemma 4.1} {\it {\rm (1)}.  The operator $A$ has only the
eigenvalues $\nu n(n + 1)(n=1,2,\ldots)$.

 {\rm (2)}.  The eigenvalue
$\nu n(n+1)$ has at most multiplicity $8(2n+1)$.}

{\bf Proof.} (1).  We can easily prove that if $\lambda$ is an
eigenvalue of $A$, then $\lambda$ has to be an eigenvalue of either
$\nu A_0$ or $\nu A_1$, or $\nu A_2$.  Then it suffices to
prove that all the eigenvalues of $A_0, A_1$ and $A_2$ are the
numbers $\{n(n+1)|n = 1,2,\ldots\}$ and count multiplicities.
According to [CH], the eigenvalues of
$A_0$ are the numbers  $\{n(n+1)|n=1,2,\ldots\}$,
 and the eigenfunctions corresponding to $n(n+1)$
 are the $2n+1$ spherical harmonics, denoted $Y_{n,h}(h
= 0,\pm 1,\ldots,\pm n)$.  Thanks to the Hodge decomposition (2.4),
we can easily prove that $\nabla Y_{n,h} (h = 0, \pm 1,\ldots,\pm n)$
are  eigenvectors of $A_1$ corresponding to the eigenvalue $n(n+1)$,
and that $\{\nabla Y_{n,h}|h = 0, \pm 1,\ldots,\pm n, n=1,2,\ldots\}$
is a complete basis of $H_1$.  Therefore, the operator $A_1$
has only eigenvalues $\{n(n+1)|n=1,2,\ldots\}$.

Similarly, we can also prove that the operator $A_2$ has only
eigenvalues $\{n(n+1)|n=1,2,\ldots\}$ with corresponding eigenvectors
$\{\nabla Y_{n,h} + \hbox{curl}\ Y_{n,h'}|h,h' = 0,\pm 1,\ldots, \pm
n\}$.

(2).  Set $\lambda = \nu n(n+1)$, and
$$
A_\lambda = A - \lambda I =
\left(\matrix
A_{0\lambda} & &B_0\cr
	&A_{1\lambda}&B_1\cr
	&&A_{2\lambda}\endmatrix\right),
$$
where $A_{i\lambda} = \nu A_i - \lambda I (i=0,1,2), B_0 = \hbox{div\
and\ }B_1 = \nabla (\hbox{div})$ and all unwritten elements of the
matrix vanish.  Then by direct computation, we can
obtain that
$$
A_\lambda^k =
\left(\matrix
A_{0\lambda}^k & &B_0^{(k)}\cr
	&A_{1\lambda}^k&B_1^{(k)}\cr
	&&A_{2\lambda}^{k}\endmatrix\right),
$$
where $B^{(k)}_i = A_{i\lambda} B^{(k-1)}_i + B_i A^{k-1}_{2\lambda},
i=0,1$.

Now we suppose that $A^k_\lambda Z = 0$ for $Z = (\zeta,v,w)$.  It
follows immediately that $A^k_{2\lambda}w = 0$ in which case $w=0$ or
$\lambda$ is an eigenvalue of $A_2$ and $w$ an associated eigenvector.
 In the first case we infer that $A^k_{0\lambda}\zeta =0, A^k_{1\lambda} v =
0$; hence $\lambda$ is one of the numbers $\nu n(n+1)$ and $\zeta$ and/or
$v$ are associated eigenvectors of $A_0, A_1$.  In the second case
$\lambda$ is again one of the numbers $\nu  n(n+1)$ and $w$ is an
associated eigenvector of $A_2$; $w = \nabla
Y_1 + \hbox{curl}\ Y_2, Y_1$ and $Y_2$ being linear combinations of the
$Y_{n,h}( h = 0,\pm 1, \ldots, \pm n)$.  We  infer that
$$\eqalign{
B^{(2)}_0w &= A_{0\lambda}B_0 w + B_0A_{2\lambda}w\cr
	&= A_{0\lambda}\hbox{div} [\nabla Y_1 + \hbox{curl}\ Y_2]\cr
	&= A_{0\lambda} \hbox{div}\ \nabla Y_1\cr
        &= A_{0\lambda} (\triangle Y_1)\cr
        &= -n (n+1)A_{0\lambda}Y_1\cr
        &=  0.\cr
}$$
Similarly, $B_1^{(2)} w=0$ and by iteration $B^{(k)}_iw =0$, for
$i=0,1$ and all $k\ge 2 $.  Hence we obtain that
$$
\hbox{ker}\ A^2_\lambda = \{(\zeta, v,w)\ |\ A_{0\lambda}\zeta = 0,
A_{1\lambda} v=0, A_{2\lambda} w=0\},
$$
and this  is the generalized eigenspace of A  with respect to $\lambda$.
In summary,
$\lambda$ has at most the multiplicity $2(2n + 1 + 2n + 1 + 2(2n+1))
= 8(2n+1)$.

The proof is complete.
\qed

To prove the existence of inertial manifold for the embedded system,
it is common to truncate the nonlinear terms $R(Z)$.  To this end, we
need

\bigskip

{\bf Lemma 4.2.}  {\it If $F\in D(A_0)$, then ${\Cal A}$ is bounded in
$D(A)$.}

\bigskip

{\bf Proof.}  It is easy to see that if $F \in D(A_0)$, then ${\Cal
A}_0$ is bounded in $D(A^{3/2}_0)$.  Then Theorem 3.2 implies  that
${\Cal A}$ is bounded in $D(A)$.
\qed

Now let $\rho$ be a positive number such that
$$
|AZ|_H\leq {\rho\over 2},\ \forall\ Z\in{\Cal A}.
$$
We consider the following prepared equations of the embedded
system:
$$
{dZ\over dt} + AZ + R_\eta (Z) = 0.\leqno(4.4)
$$
Here
$$
R_\eta (Z) = \eta_\rho (|AZ|_H) R(Z),\ \forall\ Z\in D(A),\leqno(4.5)
$$
where $\eta_\rho(s) = \eta (s/\rho)$ and  $\eta$  is a  $C^\infty$ function
from $\Bbb{R}_{+}$ into $[0,1]$ such that
$$\eta(s) = \left\{\aligned &1 \;\;\;\; \text{ for } 0\leq s\leq 1,\\
& 0 \;\;\;\; \text{ for }  s\geq 2,
\endaligned \right.
$$
and $\sup_{s\geq 0} |\eta'(s)|\leq 2$.

\bigskip

{\bf Lemma 4.3.}  {\it $R_\eta$ is a globally bounded operator from
$D(A)$ into itself:}
$$
\sup_{Z\in D(A)} |AR_\eta (Z)|_H\leq \sup_{|AZ|_H\leq 2\rho} |AR(Z)|_H
= M_1.\leqno(4.6)
$$

\bigpagebreak

{\bf Lemma 4.4}.  {\it $R_\eta$ is a globally Lipschitz operator from
$D(A)$ ito itself:}
$$
|A(R_\eta(Z_1) - R_\eta(Z_2))|_H \leq M_2|A(Z_1 - Z_2)|_H.\leqno(4.7)
$$

These lemmas are easy.  Explicit values of $M_1$ and $M_2$ are needed
and given below.

\bigskip\noindent
\subheading{\bf 4.2.  Inertial Manifolds}

Now we let
$$
0 < \lambda_1\leq \lambda_2\leq\ldots\leq
\lambda_N\leq\lambda_{N+1}\leq\ldots
$$
denote the eigenvalues of $A$ repeated according to their multipliticity.
Let $P= P_N$ be the projection associated with the eigenvalues
$\{\lambda_n|n\leq N\}$, and $Q=  Q_N = I - P$.  Then  $PH$ is the
direct sum of the eigenspaces corresponding to the eigenvalues
$\lambda_n\leq \lambda_N$, and ${ Q}H$ is the direct sum of the
eigenspaces corresponding to the eigenvalues $\lambda_n\geq
\lambda_{N+1}$.  It is easy to see that
$$
PH\subset D(A),\ \ H = PH\oplus { Q}H,\leqno(4.8)
$$
and $PH,{ Q}H$ are invariant under $A$.

The inertial manifold for equation (4.4) is obtained in the form
${\Cal M} = \hbox{Graph} \Phi$, i.e. as the graph of a Lipschitz mapping
$$
\Phi : PH\to { Q}H\cap D(A),\leqno(4.9)
$$
such that
$$
\left\{\aligned
&\hbox{supp} \Phi\subset \{y\in PH, |Ay|_H\leq 2\rho\},\\
&|A\Phi (y)|_H\leq\nu^2 b,\ \forall y\in PH,\\
&|A\Phi (y_1) - A\Phi (y_2)|_H\leq l|A(y_1-y_2)|_H,\ \forall
y_1,y_2\in PH.\endaligned
\right.
\leqno(4.10)$$

{\bf Theorem 4.1.}  {\it We can choose $N$ such that there exists an
inertial manifold for the prepared equation {(\rm 4.4)} of the
embedded system {\rm (4.1)}, which is the graph of a Lipschitz
mapping {\rm (4.9)} satisfying {\rm (4.10)}.}

{\bf Proof.}  By Lemma 4.1, it is easy to see that there is an
integer $N\geq 1$ such that
$$
\left\{\aligned &
\lambda_{N+1} - \lambda_N = 2\nu (n+1)\geq 2M_2{\displaystyle 1 + l\over l},\\
&\lambda_{N+1} = \nu(n+1)(n+2)\geq {\displaystyle M_1\over \nu^2b}.\\
\endaligned
\right.\tag4.11
$$
Then the result can be obtained by applying the theory of
inertial manifolds for nonself--adjoint operators to (4.4) (cf. [DT]
and [SY]).  The estimates (conditions) in (4.11) are those of [DT];
$\ell,b$ are arbitrarily chosen, e.g. $\ell = b = 1/2$.
\qed

Finally we have

{\bf Theorem 4.2}  {\it For $f\in D(A_0)$, the essential long--time
dynamics of the $2D$ Navier--Stokes equations on the sphere $S^2$ is
completely described by the system of ordinary differential equations
$$
{dy\over dt} + Ay = PF(y + \Phi (y)),\ \ \hbox{in}\ \ PH,\leqno(4.12)
$$
where $\Phi$ is the Lipschitz map {\rm (4.9)} obtained in Theorem 4.1
satisfying {\rm (4.10)}.}
\bigskip

\heading{ 5.  Dimension of the Inertial Manifolds}
\endheading

In order to estimate the dimension of the inertial manifold for the
embedded system given by Theorem 4.1, we need to estimate the
constants $M_1$ and $M_2$ in (4.11) and the relationship between $N$
and $n$ in (4.11).  To this end, we have
\bigskip

{\bf Lemma 5.1}.  {\it Suppose that $N$ is the largest number such
that $\lambda_N = \nu n(n+1)$, then}
$$
N\leq 8n (n+2).\leqno(5.1)
$$

{\bf Proof.}  By counting the multiplicities, we obtain that
$$
N\leq 8[(2\cdot 1 + 1) + \cdots + (2\cdot n+1)] = 8n(n+2).
$$
\qed

{\bf Lemma 5.2.}  {\it The dimension $N$ of the inertial manifold
satisfies }
$$
N\leq 1+ \max \left\{{{8M_1\over\nu^3b}, {8M_2^2(1+l)^2\over
\nu^2l^2}}\right\}.\leqno(5.2)
$$

{\bf Proof.}  By Lemma 5.1, we obtain that
$$
N\leq 8(n+1)^2,
$$
which implies that
$$
n\geq\sqrt{{N\over 8}} - 1.
$$
Therefore
$$
\lambda_{N+1} - \lambda_N = 2\nu (n+1)\geq\nu\sqrt{{N\over 2}}.
$$
Hence if we choose $N$ such that
$$N \ge  {8M^2_2(1+l)^2\over \nu^2l^2},
$$
then (4.11)$_1$ is satisfied.  Moreover, it is easy to see that if we
choose $N$ such that
$$
N \ge  {8M_1\over \nu^3b},
$$
then (4.11)$_2$ is satisfied.

The proof is complete.
\qed

To estimate $M_1$ and $M_2$ we need to estimate $\rho$ as follows.

\bigskip

{\bf Lemma 5.3}.
$$
|AZ|\leq\rho = c\nu^2[G_0G_4 + G^{12}_0],\ \forall\ Z\in {\Cal
A},\leqno(5.3)
$$
{\it where $G_0$ and $G_4$ are generalized Grashof numbers\footnote{No length
appears in the definition (5.4) of $G_0$  and $G_4$  because we assumed
that the radius $a$  of the sphere is equal to one. For $a\ne 1$, we would
define $G_0$  and $G_4$  by
$$G_0=\frac{|f|a^2}{\nu^2},\ \ G_4 = {\|f\|_{H^4}a^6 \over \nu^2}.$$}
defined
by}
$$
G_0 = {|f|\over\nu^2},\ \ G_4 = {\|f\|_{H^4}\over \nu^2}.\leqno(5.4)
$$

{\bf Proof.}  {\it Step 1.}  We estimate the bound for
$\zeta\in{\Cal A}_0$ in $D(A^{3/4}_0$).  To this end, we have from
(1.8) that
$$
{1\over 2} {d\over dt}|\zeta|^2 + \nu|A^{1/2}_0\zeta|^2 =
(F,\zeta)\leq {c|f|^2\over \nu} + {\nu \over 2}|A^{1/2}_0\zeta|^2,
$$
which implies that
$$
{d\over dt}|\zeta|^2 + \nu|A^{1/2}_0\zeta|^2\leq
{c|f|^2\over \nu}.\leqno(5.5)
$$
By Gronwall's  inequality, we obtain that
$$
|\zeta (t)|^2 \leq {c|f|^2\over \nu^2},\leqno(5.6)
$$
for $t$ large enough.  Since the attractor ${\Cal A}_0$ is time invariant,
by shifting the time $t$, (5.6) holds true for any $t\geq 0$.

Now we derive as in [T3] (see Sec. 4) the higher order norm estimates:
$$\eqalign{
{1\over 2}& {d\over dt} |A^{3/2}_0\zeta|^2 + \nu|A^2_0\zeta|^2\cr
&= (F,A^3_0\zeta) - (\nabla\zeta\cdot u, A^3_0\zeta)\cr
&\leq |A_0F|\cdot |A^2_0\zeta| + |(D^2(\nabla\zeta)\cdot u +
	D(\nabla\zeta)\cdot D u + D(\nabla\zeta)\cdot D^2u,
	A^2_0\zeta)|\cr
& \leq |A^2_0\zeta| \cdot\left\{\|f\|_{H^3} + \|\zeta\|_{H^{7/2}}\cdot|\zeta| +
	\|\zeta\|_{H^{5/2}}\cdot\|\zeta\|_{H^{1/2}} +
	\|\zeta\|^2_{H^{3/2}}\right\}\cr
& \leq |A^2_0\zeta| \cdot\left\{\|f\|_{H^3} + |A^2_0\zeta|^{7/8}
	\cdot |\zeta|^{9/8} +
	|A^2_0\zeta|^{6/8}\cdot  |\zeta |^{10/8}\right\}\cr
&\leq {\nu\over 2}|A^2_0\zeta|^2 + {c\|f\|^2_{H^3}\over\nu} +
{c|\zeta|^{18}\over\nu^{15}},\cr}
$$
i.e.
$$
{d\over dt}|A^{3/2}_0\zeta|^2 + \nu|A^2_0\zeta|^2 \leq
{c\|f\|^2_{H^3}\over \nu} + {c|\zeta|^{18}\over \nu^{15}}\leq c\nu^3
[G^2_3 + G^{18}_0],\leqno(5.7)
$$
where $G_3$ is the generalized Grashof number defined by
$$
G_3 = {\|f\|_{H^3}\over\nu^3}\; ( \leq CG^{1/4}_0 G^{3/4}_4\quad \hbox{ by
interpolation)} .\leqno(5.8)
$$

Using Gronwall's inequality, we can show that
$$
|A^{3/2}_0\zeta(t)|^2\leq c\nu^2[G^2_3 + G^{18}_0],\ \ \hbox{for}\ \
t\ \hbox{large enough}.
$$
Therefore
$$
|A^{3/2}_0\zeta|^2\leq c\nu^2[G^2_3 + G^{18}_0],\ \forall\ \zeta\in
{\Cal A}_0.\leqno(5.9)
$$

{\it Step 2.}  We estimate the bounds of ${\Cal A}_0$ in
$D(A^{1/2}_0)$ and $D(A_0)$.
$$\eqalign{
{1\over 2} &{d\over dt}|A^{1/2}_0\zeta|^2+ \nu|A_0\zeta|^2\cr
&= (F,A_0\zeta) - (\nabla\zeta\cdot u, A_0\zeta)\cr
&\leq c|A_0\zeta|\cdot\{\|f\|_{H^1} +
	\|\zeta\|_{H^{3/2}}\cdot|\zeta|\}\cr
&\leq c|A_0\zeta|\cdot \|f\|_{H^1} +
	c|A_0\zeta|^{7/4}\cdot|\zeta|^{5/4}\cr
&\leq {\nu\over 2}|A_0\zeta|^2 + {c\|f\|^2_{H^1}\over\nu} +
{c|\zeta|^{10}\over\nu^7} ,\cr
}$$
which implies that
$$\eqalign{
{d\over dt} |A^{1/2}_0\zeta|^2 + \nu|A_0\zeta|^2 &\leq c\nu^3
	\left[{\|f\|^2_{H^1}\over \nu^4} + G^{10}_0\right]\cr
&\leq c\nu^3\left[G^{4/3}_0 G^{2/3}_3 + G^{10}_0\right].\cr
}$$
It follows that
$$
|A^{1/2}_0\zeta (t)|^2 \leq c\nu^2 \left[G^{4/3}_0 G^{2/3}_3 +
G^{10}_0\right],\ \ \hbox{for\ $t$\ large}.
$$
Therefore
$$
|A^{1/2}_0\zeta|^2 \leq c\nu^2 \left[G^{4/3}_0 G^{2/3}_3 +
G^{10}_0\right],\ \forall\ \zeta\in{\Cal A}_0.\leqno(5.10)
$$
Moreover, we also have
$$\eqalign{
{1\over 2}&{d\over dt}|A_0\zeta|^2 + \nu|A^{3/2}_0\zeta|^2\cr
&\leq |A^{1/2}_0F|\cdot |A^{3/2}_0\zeta| + |(D^2\zeta\cdot u +
	\nabla\zeta\cdot D u, A^{3/2}_0\zeta)\cr
&\leq |A^{3/2}_0\zeta|\cdot\{\|f\|_{H^2} + \|\zeta\|_{H^{5/2}}\cdot
	|\zeta| + \|\zeta\|_{H^{3/2}}\cdot\|\zeta\|_{H^{1/2}}\}\cr
&\leq |A^{3/2}_0\zeta|\cdot\left\{\|f\|_{H^2} + |A^2_0\zeta |^{6/5}\cdot
	|\zeta|^{7/6}\right\}\cr
&\leq {\nu\over 2} |A^{3/2}_0\zeta|^2 + {c\|f\|^2_{H^2}\over\nu} +
	{c|\zeta|^{14}\over \nu^{11}}.\cr
}$$
Therefore, we obtain that
$$
{d\over dt} |A_0\zeta|^2 + \nu|A^{3/2}_0\zeta|^2\leq c\nu^3
\left[G^{2/3}_0 G^{4/3}_3 + G^{14}_0\right].
$$
Hence by Gronwall's inequality, we obtain as before that
$$
|A_0\zeta|^2\leq c\nu^2\left[G^{2/3}_0 G^{4/3}_3 + G^{14}_0\right],\
\forall\ \zeta\in {\Cal A}_0.\leqno(5.11)
$$

{\it Step 3.}  Now we estimate the bound for ${\Cal A} = J({\Cal
A}_0)$.  To this end, for any $Z = (\zeta,v,w)\in {\Cal A}$, we have
$v = \nabla\zeta$ and $w = \zeta\ \hbox{curl}\ (A^{-1}_0\zeta)$.  It follows
that
$$\leqalignno{
|A_1v|^2 &\leq c\nu^2[G^2_3 + G^{18}_0],&(5.12)\cr
|A_2w|^2 &\leq c\|\zeta\|^2_{H^2}\cdot\|\zeta\|^2_{H^1}&(5.13)\cr
&\leq (\hbox{by}\ (5.10) - (5.11))\cr
&\leq c\nu^4 [G^{2/3}_0 G^{4/3}_3 + G^{14}_0]\cdot
	[G^{4/3}_0 G^{2/3}_3 + G^{10}_0].\cr
}$$
Then we can easily obtain using (5.8) that
$$
\nu^2\{|A_0\zeta|^2 + |A_1\nu|^2 + \alpha|A_2w|^2\}\leq c\nu^4[G_0G_4
+ G^{12}_0]^2,
$$
which shows that
$$
|AZ|_H\leq c\nu^2 [G_0G_4 + G^{12}_0],\ \forall\ Z\in {\Cal
A}.\leqno(5.14)
$$
The proof is complete.
\qed

Now we need to estimate $M_1$ and $M_2$.

{\bf Lemma 5.4}.
$$
M_1\leq c\nu^3[G^6_0 G^6_4 + G^{72}_0].\leqno(5.15)
$$

{\bf Proof.}  For any $Z = (\zeta,v,w)\in D(A)$ such that $|AZ|_H\leq
2\rho$, we have
$$\eqalign{
|AR(Z)|^2_H &\leq 2 \nu^2\{|A_0R_0(Z)|^2 + |A_1R_1(Z)|^2 +
	\alpha|A_2R_2(Z)|^2\}\cr
&\qquad + 2 |\hbox{div}\ R_2(Z)|^2 +2 |\nabla\ \hbox{div}\ R_2(Z)|^2\cr
&\leq 2 \nu^2\{\|f\|^2_{H^3} + \|f\|^2_{H^4}\} + c|A_2R_2(Z)|^2\cr
&\leq 2 \nu^2\{\|f\|^2_{H^3} + \|f\|^2_{H^4}\} + c
	 \{\|f\|^2_{H^3} \|\zeta\|^2_{H^1}+ \|f\|^2_{H^2}
	 \|\zeta\|^2_{H^2} + \|v\|^2_{H^2}\|\zeta\|^4_{H^2}\cr
&\qquad + \nu^2\|v\|^2_{H^2}  \|\zeta\|^2_{H^2} + \nu^2(1 + \nu^{-4}
	  \|\zeta\|^4_{H^{1/2}})^2 \|w - \tilde w\|^2_{H^2}\}\cr
&\leq 2 \nu^2\{\|f\|^2_{H^3} + \|f\|^2_{H^4}\} + c
	\Big\{\|f\|^2_{H^3}{\rho^2\over \nu^2} +
	 \|f\|^2_{H^2}{\rho^2\over\nu^2} + {\rho^6\over\nu^6}\cr
&\qquad + \nu^2(1 + v^{-8}\rho^4)^2(\rho^2 +
	{\rho^4\over\nu^4})\Big\}\cr
& \le (\text{with  } (5.3) ) \cr
&\leq c\nu^6 [G^6_0G^6_4 + G^{72}_0]^2.\cr
}$$
Therefore, by (4.6) we obtain (5.15).
\qed

{\bf Lemma 5.5.}
$$
M_2\leq c\nu[G^5_0G^5_0 + G^{60}_0].\leqno(5.16)
$$

{\bf Proof.}  As shown in [T1] (see (2.19) in Chapter VIII), $M_2$ is given by
$$
M_2 = {2M_1\over \rho} + c_{2\rho},\leqno(5.17)
$$
where $c_{2\rho}$ is the local Lipschitz constant of $R$ defined by
$$
|A(R(Z_1) - R(Z_2))|_H\leq c_{2\rho}|A(Z_1 - Z_2)|_H,\ \forall\
Z_i\in D(A), |AZ_i|_H\leq 2\rho,i=1,2.\leqno(5.18)
$$

Obviously,
$$
{2M_1\over \rho}\leq c\nu [G^5_0G^5_4 + G^{60}_0].\leqno(5.19)
$$

Now for any $Z_i \in D(A),\ |AZ_i|_H\leq 2\rho, i=1,2$, we have
$$\eqalign{
|A(R(Z_1) - R(Z_2))|_H &\le  |\hbox{div}\ (R_2(Z_1) - R_2(Z_2))|\cr
&\qquad + |\nabla\ \hbox{div}\ (R_2(Z_1) - R_2(Z_2))| + |A_2(R_2(Z_1)
	- R_2(Z_2))|\cr
&\le (\text{since  } A_2: D(A_2)\to H_2 \text{  is an isomorphism })\cr
&\leq c|A_2(R_2(Z_1) - R_2(Z_2))|\cr
&\leq |A(Z_1 - Z_2)|_H\cdot \left[{\|f\|_{H^3}\over \nu} +
	{\rho^3\over \nu^3} + {\rho\over\nu}\right]\cr
&\qquad + c\nu(1 + \nu^{-8}\rho^4)|A_2(w_1 - \tilde w_1 - (w_2 - \tilde
	w_2))|\cr
&\qquad + c\nu^{-3}(|A_2 w_1| + |A_2\tilde
	w_1|)(\|\zeta_1\|^4_{H^{1/2}} - \|\zeta_2\|^4_{H^{1/2}})\cr
&\leq |A(Z_1 - Z_2)|_H \cdot\Big[\nu G_3 + {\rho^2\over \nu^3} +
	c\nu (1 + \nu^{-8}\rho^4)\cr
&\qquad\qquad +\nu^{-1} (1 + \nu^{-8}\rho^4)\rho + {\rho^4\over\nu^7} +
	{\rho^5\over\nu^9}\Big].\cr
}$$

Hence
$$\leqalignno{
c_{2\rho} &\leq \left[\nu G_3 + {\rho^2\over\nu^3} + c\nu(1 +
	\nu^{-8}\rho^4) +\nu^{-1}(1 - \nu^{-8}\rho^4)\rho +
	{\rho^4\over \nu^7} + {\rho^5\over\nu^9}\right]&(5.20)\cr
&\leq c\nu[G^5_0 G^5_0 + G^{60}_0].\cr
}$$

Then the combination of (5.17) and (5.19)--(5.20) proves (5.16).
\qed

We are in position to state the main theorem in this section,
which is now a direct consequence of Lemmas 5.2 and 5.4 -- 5.5.

\bigskip

{\bf Theorem 5.1.}  {\it The dimension $N$ of the inertial manifold
given by Theorem 4.1 is bounded as follows:}
$$
N\leq c[G^5_0G^5_4 + G^{60}_0]^2.\leqno(5.21)
$$

\bigpagebreak

{\bf Remark 5.1.} Let us mention here the result in [EFNT]
concerning exponential attractors. Exponential attractors are (nonsmooth)
positively invariant sets that attract all orbits  at an exponential rate.
For general dissipative equations, and in particular for the 2D Navier-Stokes
equations with general boundary conditions,
existence of such sets was shown in [EFNT]; furthermore the upper estimate
of their dimension
can be the same as that of the attractor, i.e. $cG_0$.
For the equation we consider, the dimension of the
exponential attractor is much smaller than that of the inertial manifold;
however it is not a smooth manifold.

\qed
\bigskip
\heading{6.  Complementary Results}
\endheading
\bigskip\noindent

\subheading{6.1.  Remarks on the Space Periodic Case}

We briefly present here a simplified proof of the result
 of [K1]  on the existence of inertial forms
for the 2D Navier-Stokes equations with periodic boundary conditions
when the ratio of the periods  is rational. This proof differs
from [K1] in that the embedded reaction-diffusion system is much simpler
and the analysis is consequently simplified.

Let $x = (x^1,x^2)$ be the coordinate for the $2D$ torus $\bbT^2 =
\bbR^2/(2\pi\bbZ)^2$.  We consider the $2D$ Navier--Stokes equations
on $\bbT^2$:
$$
\left\{\aligned &u_t + (u\cdot\nabla)u + \nabla p - \nu\Delta u = f,\\
&\hbox{div}\ u = 0,\\
&u\big|_{t=0} = u_0,\\
\endaligned \right. \tag6.1
$$
where $u = u(x,t) = (u^1,u^2)$ is the velocity field and $p = p(x,t)$
is the pressure.

We  define  as usual  the following function spaces (cf. [T1]):
$$
\left\{\aligned
&H_0 = \{u\in L^2(\bbT^2)^2|{\displaystyle\int_{\bbT^2}} udx^1 dx^2 = 0,
\text{div } u=0,\}\\
& V_0 = H_0\cap H^1(\bbT^2)^2,\\
&H_1 = (H_0)^2, H_2 = L^2(\bbT^2)^4,\\
&V_1 = (V_0)^2, V_2 = H^1(\bbT^2)^4,\\
&H = H_0\times H_1\times H_2,\\
&V = V_0\times V_1\times V_2.\endaligned\right. \leqno(6.2)
$$
We also consider  the orthogonal projectors:
$$
\left\{\aligned
&P_0 :L^2(\bbT^2)^2\to H_0,\\
&P_1 : L^2(\bbT^2)^4\to H_1,\\
&P : L^2(\bbT^2)^8\to H.
\endaligned \right. \tag6.3
$$

Now we define an injection map $J : V_0\to H$ by  setting
$$
J(u) = (u,\tilde v,\tilde w),\leqno(6.4)
$$
where
$$
\left\{\aligned
&\tilde v = \nabla u = u^i_{;j}e_i\otimes e^j,\\
&\tilde w = u\otimes u = u^i u^j e_i\otimes e_j,\\
\endaligned \right.
$$
$e_i (i=1,2)$ being the unit vectors in $x^1$ and $x^2$ directions
respectively, with dual basis denoted by $\{e^1,e^2\}$.

Then we can consider the following embedded system
$$\left\{\aligned &
u_t = \nu A_0u + P_0 (w^{ik}_{;k} e_i) = f,\\
&v_t + \nu A_1v + P_1 [w^{ik}_{;kj}e_i\otimes e_j] = \nabla f,\\
 & w_t + \nu A_2w + u\otimes P_0 Tr (u\otimes v) + P_0 Tr(u\otimes
	v)\otimes u\\
&\qquad + 2\nu\sum^2_{k=1} v^i_k v^j_k e_i\otimes e_j -
	(f\otimes u + u\otimes f)\\
&\qquad + \underline{k\nu [1 + \nu^{-4}\|u\|^4_{H^{3/4}} + \nu^{-2}
	(\|u\|^2_{H^{3/2}} + \|v\|^2_{H^{1/2}})](w-u\otimes u)} =
	0.\endaligned \right.\leqno(6.5)
$$

It is easy to see that $\{ u, v=\tilde v, w=\tilde w\}$   satisfy
equations (6.5), the underlined terms in (6.5) being equal to 0.
Considering then the system (6.5), we can prove
exactly as we did for the NSE on the sphere $S^2$,
that for $k$ large
enough, (6.5) is a dissipative system.  Then all results in [K1] can
be proved.  Especially, there is a global attractor ${\Cal A}$ for
the embedded system (6.5) given by
$$
{\Cal A} = J({\Cal A}_0),\leqno(6.6)
$$
${\Cal A}_0$ being the global attractor of the original
Navier--Stokes equations (6.1).  Moreover, by studying the prepared
system of (6.5) as (4.4) one  can prove the existence of inertial
manifolds for the prepared system of (6.5).  As a consequence, there
is an inertial form of (6.1), which reproduces all the dynamics of the
Navier--Stokes equations (6.1).

\newpage

\bigskip\noindent
\subheading{6.2.  The Barotropic Equations of the Atmosphere}

All our previous results are also true after some modifications for
the following $2D$ barotropic vorticity equation of the  globe atmosphere,
which was derived by Rossby in the  1930s, and was used by J.G. Charney,
R. Fj\o rtoft \& J. von Neumann [CFN] in the first weather predictions
in the  1950s:
$$
\left\{\aligned
&u_t + \nabla_u u + 2 \Omega \cos \theta k\times u -\nu \triangle u = f, \\
& u\big|_{t=0} = u_0,\endaligned \right. \tag6.7
$$
where the space domain is $S^2_a$, the 2D sphere with radius a,
$\Delta$ is the Laplace--Beltrami
operator on $S^2_a$, k is the unit outward normal vector, $\Omega$
is the angular velocity of the earth, and  $\theta$  is the
colatitude of the earth. The
existence and dimension estimates of
attractors of these equations (in vorticity form) were studied in [W].
We remark here
that all the results in Sections 1--5 hold true for equations (6.7).
Particularly, we can obtain an inertial form of (6.7) with dimension
given by an estimate similar to (5.21).
\vskip.5truein
\heading{Acknowledgement}
\endheading
\bigskip

This work was partially supported by the National Science Foundation
under Grant NSF-DMS-9024769, by Department of Energy under Grant
DE-FG02-92ER25120, and by the Research Found of Indiana University.

\bigpagebreak

\end